\documentclass[preprint,12pt]{elsarticle}

\usepackage{amsmath}
\usepackage{amssymb} 
\usepackage{bm}

\newcommand{\fourier}[1]{\mathcal{F} \biggl[#1 \biggl]}
\newcommand{\ifourier}[1]{\mathcal{F}^{-1} \biggl[#1 \biggl]}
\newcommand{\sfourier}[1]{\mathcal{F} \bigl[#1 \bigr]}
\newcommand{\sifourier}[1]{\mathcal{F}^{-1} \bigl[#1 \bigr]}

\newcommand{\x}[1]{{\bf x}_{#1}}

\newcommand{\sprop}[3]{\mathcal{{\rm Prop}}^{#1}_{#2} \bigl[#3\bigl]}

\newcommand{\rot}[2]{{\rm Rot}_{\rm #1} \bigl[#2\bigl]}

\newcommand{\eq}[1]{Eq.(\ref{#1})}

\journal{}

\begin{document}

\begin{frontmatter}



\title{Numerical investigation of lensless zoomable holographic multiple projections to tilted planes}


\author[a]{Tomoyoshi Shimobaba\corref{cor1}}
\cortext[cor1]{Tel: +81 43 290 3361; fax: +81 43 290 3361}
\ead{shimobaba@faculty.chiba-u.jp}
\author[b]{Michal Makowski}
\author[a]{Takashi Kakue}
\author[a]{Naohisa Okada}
\author[a]{Yutaka Endo}
\author[a]{Ryuji Hirayama}
\author[a]{Daisuke Hiyama}
\author[a]{Satoki Hasegawa}
\author[a]{Yuki Nagahama}
\author[a]{Tomoyoshi Ito}

\address[a]{Chiba University, Graduate School of Engineering, 1--33 Yayoi--cho, Inage--ku, Chiba, Japan, 263--8522}
\address[b]{Faculty of Physics, Warsaw University of Technology, 75 Koszykowa, 00-662 Warsaw, Poland}

\begin{abstract}
This paper numerically investigates the feasibility of lensless zoomable holographic multiple projections to tilted planes.
We have already developed lensless zoomable holographic single projection using scaled diffraction, which calculates diffraction between parallel planes with different sampling pitches.
The structure of this zoomable holographic projection is very simple because it does not need a lens; however, it only projects a single image to a plane parallel to the hologram.
The lensless zoomable holographic projection in this paper is capable of projecting multiple images onto tilted planes simultaneously.
\end{abstract}

\begin{keyword}
Computer-generated hologram \sep Fresnel diffraction \sep Holographic projection \sep Scaled diffraction \sep Tilted diffraction

\end{keyword}

\end{frontmatter}

\section{Introduction}
Holographic projection \cite{proj1,specklegs1,specklegs2} is a kind of laser projection.
It requires two-step processing for the projection.
The first step is to calculate a hologram from the image to be projected using diffraction calculation.
In the second step, a spatial light modulator (SLM) displaying the hologram reconstructs the projected image.
The optical system does not need a lens because the hologram itself has the property of a lens. 
Therefore, the system, which ultimately requires only an SLM and light source, promises to be very simple.
A holographic projection is especially suitable for micro- and pico-projectors due to the above-mentioned reasons, but there are problems: for example, speckle noise, long calculation time for hologram, and magnification of a projected image. 

Speckle noise is a common problem on laser projections.
In holographic projection, many methods for speckle noise reduction have been proposed.
Iterative algorithms based on Gerchberg-Saxton (GS) algorithm \cite{gs} are widely used to gradually improve the speckle noise by iterating diffraction and inverse diffraction calculations with constraints on the reconstructed and hologram planes by {\it a priori} information.
Speckle noise reduction by sparsely placing the pixels on a projected image has been proposed \cite{takaki}. 
Following which, a pixel separation method \cite{pixel} has been proposed to improve the reducing resolution of the projected image using the speckle noise reduction method of Ref. \cite{takaki}.

The problem of the hologram calculation time is a critical issue in holographic projection because multiple diffraction calculations are required to efficiently reduce the speckle noise.
For example, fast hologram calculation algorithms have been proposed \cite{alg1,alg2}.
Hardware-based methods using graphics processing units (GPUs) and field programmable gate arrays (FPGAs) have also been proposed \cite{cwo,horn5}. 

A zoom function is indispensable to projectors.
A zoom lens module consisting of mechanical parts and many lenses to avoid aberrations is used in general, but using it increases the system size, weight and cost.
In addition, it requires manual operation of the module to adjust to the proper magnification.
In holographic projection, a liquid crystal (LC) lens, which can magnify the projected image by displaying a Fresnel lens pattern on it, was demonstrated instead of using a zoom lens \cite{lc,lc2}.
The method could control the magnification by the Fresnel lens pattern without manual operation.
However, this method requires an additional special device. 
Whereas, we proposed a zoomable holographic projection without any lens or special device \cite{zoom1,zoom2}.
The method realized the zoom function using scaled diffraction \cite{scale1,scale2,scale3,scale4,scale5,scale6,scale7,scale8,scale9}, which calculates diffraction between planes with different sampling pitches.
The magnification is determined by the ratio of the sampling pitch on the projected image and that of the hologram.
The structure of this zoomable holographic projection is very simple because it does not need a lens; however, it only projects a single image to a plane parallel to the hologram.

This paper numerically investigates the feasibility of lensless zoomable holographic multiple projections to tilted planes.
The lensless zoomable holographic projection in this paper is capable of projecting multiple images onto multiple tilted planes simultaneously.

\section{Lensless zoomable holographic projection to multiple tilted planes}

\begin{figure}[htb]
\centerline{\includegraphics[width=12cm]{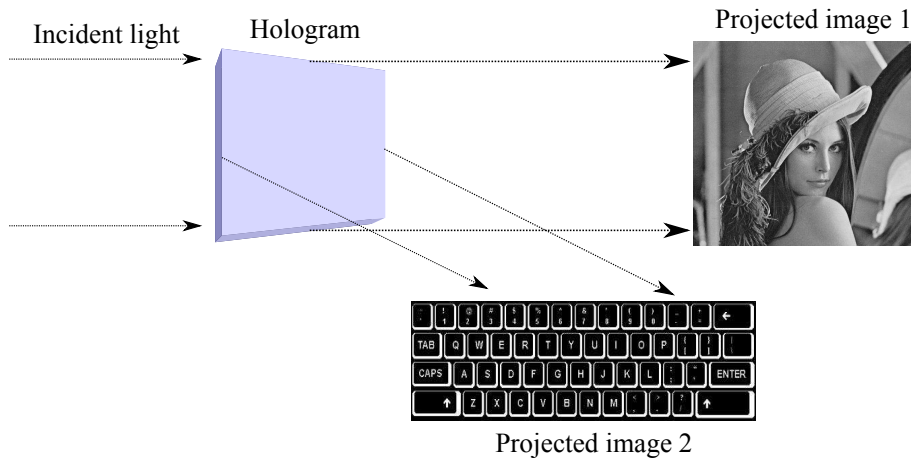}}
\caption{Our lensless zoomable holographic projection to tilted planes.}
\label{fig:system}
\end{figure}

Reference \cite{tilt_prj} has already reported a holographic projection to a tilted plane using fractional Fourier transform with a lens; however, it does not consider multiple projections to tilted planes, and could not magnify the projected images.
In contrast, as shown in Fig. \ref{fig:system}, our lensless zoomable holographic projection is considered multiple projections to tilted planes.
Our system does not include a lens.
Here, we assume that the system can simultaneously project two images: ``Projected Image 1'' is parallel to the hologram and ``Projected Image 2'' is off-axis and non-parallel to the hologram.
In addition, each projected image is individually magnified without a zoom lens.

The application of this projection system is that, for example, the system projects the image of the keyboard (``Projected Image 2'') on a desk and  another image (``Projected Image 1'') on a screen from one SLM.
If we had sufficient computational power for real-time hologram calculation, we would be able to interactively operate Projected Image 1 on the screen by pushing the keyboard image.
Pushing the keyboard image would be detected by a camera and computer vision algorithms.
We do not mention the real-time hologram calculation hereafter.

In our previous work \cite{zoom1, zoom2}, we magnified a single projected image parallel to the hologram using scaled diffraction, but could not realize a 
projected image on a tilted plane.
This paper extends scaled diffraction so that it can project onto a tilted screen.
Let us begin with scaled diffraction.
Many scaled diffraction calculations have been proposed \cite{scale1,scale2,scale3,scale4,scale5,scale6,scale7,scale8,scale9}.
Scaled diffraction is commonly expressed by,
\begin{equation}
u_h(\bm x_h) = \sprop{p_1,p_h}{z}{u_1(\bm x_1) },
\label{eqn:scale}
\end{equation}
where the operator $\sprop{p_1,p_h}{z}{u_1(\bm x_1)}$ indicates scaled diffraction, $u_1(\bm x_1)$ is the source plane (an image to be projected) with the pixel pitch of $p_1$, $u_h(\bm x_h)$ is the complex amplitude on a hologram with the pixel pitch of $p_h$, and $z$ is the propagation distance between the source and destination planes.
$\bm x_1=(x_1,y_1)$ and $\bm x_h=(x_h,y_h)$ mean the position vector on the source and hologram.
The planes $u_1(\bm x_1)$ and $u_h(\bm x_h)$ are placed in parallel to each other.
There are some implementations of scaled diffraction: for example, shifted Fresnel diffraction \cite{scale3}, ARSS Fresnel diffraction \cite{scale9}, scaled angular spectrum method \cite{scale8, nu} and so forth.

Here, we use ARSS Fresnel diffraction as the scaled diffraction.
ARSS Fresnel diffraction improves the problem of shifted Fresnel diffraction for aliasing noise.
In one-dimension, it is expressed by, 
\begin{eqnarray}
u_h(x_h) &=& C_z \ifourier{ 
\fourier{u_1(x_1) \exp(i \phi_u)} 
\fourier{\exp(i \phi_h) {\rm Rect}(\frac{x_h}{2 x_{max}}) } },
\label{eqn:arss}
\end{eqnarray}
where ${\rm Rect}(\cdot)$ is the rectangular function for the band-limitation to $\exp(i \phi_h)$ and $\exp(i \phi_u)$, $\exp(i \phi_h)$ and $C_z$ are defined by,
\begin{eqnarray}
&&\exp(i \phi_u)= \exp(i \pi \frac{(s^2-s)x_1^2-2s o_x x_1}{\lambda z}), \\
&&\exp(i \phi_h)= \exp(i \pi \frac{s x_h^2}{\lambda z}), \\
&&C_z=\frac{\exp(i \phi_c)}{ i \lambda z }=\frac{\exp(i kz + \frac{i \pi}{\lambda z}((1-s)x_2^2+2 o_x x_2+o_x^2))}{i \lambda z} ,
\label{eqn:cz}
\end{eqnarray}
where $\lambda$ and $k$ are the wavelength and wavenumber and we define the scale parameter $s=p_1/p_h$ and the offset vector ${\bm o}=(o_x, o_y)$ away from the origin. 
The expansion to the two-dimension is straightforward.

We can convert the complex amplitude $u_h(\bm x_h)$ to the amplitude hologram $I(\bm x_h)$ by, e.g. $I(\bm x_h)=\Re(u_h(\bm x_h))$, and convert one to the kinoform $\psi(\bm x_h)$ by $\psi(\bm x_h)={\rm arg}(u_h(\bm x_h))$ where ${\rm arg}(\cdot)$ takes the argument of the complex number.
If we set $p_h$ larger than $p_1$, we can obtain a magnified image on a projected plane without using a lens .

However, \eq{eqn:scale} does not realize the projection of an image on a tilted plane.
In order to improve this issue, we use scaled diffraction with tilted diffraction \cite{tilt1,tilt2,tilt3,tilt4,tilt5}, which calculates diffraction between non-parallel planes. 
Tilted diffraction calculation of the rotation of a complex amplitude in the frequency domain with rotation matrix $\rm M$ where,
\begin{equation}
\rm M =
\begin{bmatrix}
a_{11} & a_{12} & a_{13} \\
a_{21} & a_{22} & b_{23} \\
a_{31} & a_{32} & a_{33} 
\end{bmatrix}.
\end{equation}
The relation between the source plane $U_1(u,v)=\sfourier{u_1(x_1,y_1)}$ and the rotated plane $U_1'(u',v')=\sfourier{u_1(x_1',y_1')}$ on the frequency domain is expressed by,
\begin{equation}
U_1(u,v)=\rot{M}{U_1'(u',v')}=U_1'(\alpha(u,v)-1/\lambda , \beta(u,v)-1/\lambda),
\label{eqn:rot}
\end{equation}
where $\alpha(u,v)=a_{11} u + a_{12} v + a_{13} w(u,v)$, $\beta(u,v)=a_{21} u + a_{22} v + a_{23} w(u,v)$, and $w(u,v)=\sqrt{\lambda^{-2}-u^2-v^2}$.
\eq{eqn:rot} requires the interpolation. 
We used linear interpolation as the interpolation.

We can simply obtain the scaled diffraction on a tilted plane by the combination of \eq{eqn:scale} and \eq{eqn:rot}:
\begin{equation}
u_h(\bm x_h) = \sprop{p_1,p_h}{z, \rm M}{u_1(\bm x_1)} = {\sprop{p_1,p_h}{z}{\sifourier{\rot{M}{ \sfourier{ u_1(\bm x_1) } } }}} .
\label{eqn:rot_scale}
\end{equation}
We can calculate a hologram that is capable of projecting zoomable multiple images to tilted planes by,
\begin{equation}
u_h(\bm x_h) = \sum_{j=1}^{N} \sprop{p_j, p_h}{z, \rm M_j}{u_j(\bm x_j)},
\label{eqn:multi_rot_scale}
\end{equation}
where $N$ is the number of the multiple images, $u_j(\bm x_j)$ and $\bm M_j$ are $j$-th images and the corresponding rotation matrix, respectively.

\section{Results}
\begin{figure}[htb]
\centerline{\includegraphics[width=12cm]{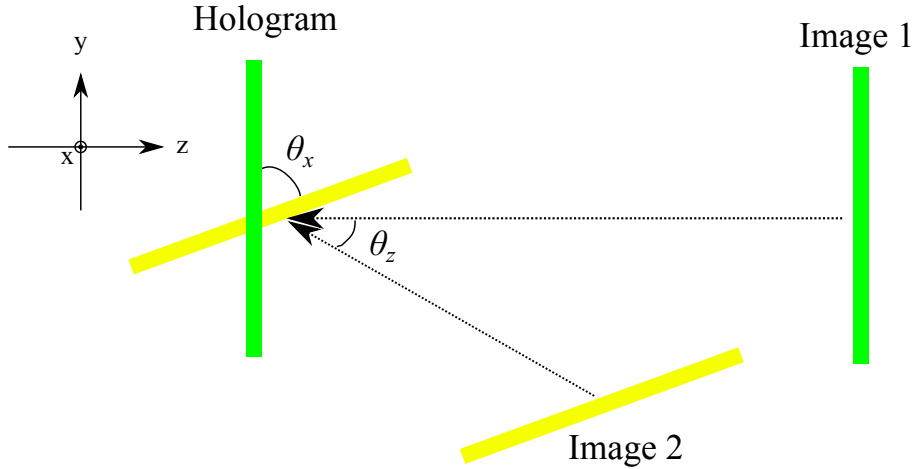}}
\caption{Setup for the hologram calculation and reconstruction.}
\label{fig:fig_calc}
\end{figure}

We show multiple reconstructed images from a hologram generated by a combination of the scaled diffraction and tilted diffraction.
The hologram is in fact calculated as kinoform.
The setup for the hologram calculation and reconstruction is shown in Fig.\ref{fig:fig_calc}.
We assume that the wavelength of light is $\lambda=633$nm and the sampling pitch on the hologram is $p_h=8 \mu$m.
We use two images with $2,048 \times 2,048$ pixels to be projected that are the same as ``Projected Image 1'' and ``Projected Image 2'' in Fig.\ref{fig:system}.
Image 1, which is denoted as $f_1(\bm x_1)$, is parallel to the hologram and the distance of $z_1=1.5$m from the hologram, whereas Image 2 which is denoted as $f_2(\bm x_2)$ is inclined to x-axis at $\theta_x=60^{\circ}$ degrees and the distance of $z_2=0.5$m from the hologram.
We use the following rotation matrix,
\begin{equation}
\bm M =
\begin{bmatrix}
1 & 0 & 0 \\
0 & \cos \theta_x & \sin \theta_x \\
0 & -\sin \theta_x & \cos \theta_x
\end{bmatrix}.
\end{equation}

The complex amplitude $u_1(\bm x_1)$ on Image 1 with random phase is expressed by,
\begin{equation}
u_1(\bm x_1) = f_1(\bm x_1) \exp(2 \pi i n(\bm x_1)), 
\end{equation}
where $n_1(\bm x_1) \in [0,1)$ means random numbers.
The diffracted result of $u_1(\bm x_1)$ on the hologram is calculated by ARSS Fresnel diffraction of \eq{eqn:arss} with a different sampling pitch.
The complex amplitude $u_2(\bm x_2)$ on Image 2 with a different random phase is expressed by,
\begin{equation}
u_2(\bm x_2) = f_2(\bm x_2) \exp(2 \pi i n(\bm x_2)) \exp(i ky \sin \theta_z), 
\end{equation}
where $n_2(\bm x_2) \in [0,1)$ also means random numbers.
The last term means that Image 2 is placed off-axis and it travels to the hologram at the angle $\theta_z$.
In addition, we use the offset $\bm o=(0,z_2 \tan(\theta_z))$ in ARSS Fresnel diffraction.
We assume $\theta_z=2.2^{\circ}$ because the sampling pitch of hologram $p_h=8 \mu$m, so that we need to set the angle within the maximum angle $\sin^{-1}(\lambda / 2p_h) \approx 2.3^{\circ}$.

The complex amplitude of $u_2(\bm x_2)$ is first rotated by the tilted diffraction with the angle $\theta_x=60^{\circ}$ and then propagated by ARSS Fresnel diffraction with different sampling pitches.
The hologram is generated by taking the argument of accumulating these complex amplitudes.
In the reconstruction, Image 1 is reconstructed simply by using ARSS Fresnel diffraction, while Image 2 is reconstructed by firstly calculating ARSS Fresnel diffraction and then rotating the diffracted result by \eq{eqn:rot}.

\begin{figure}[htb]
\centerline{\includegraphics[width=14cm]{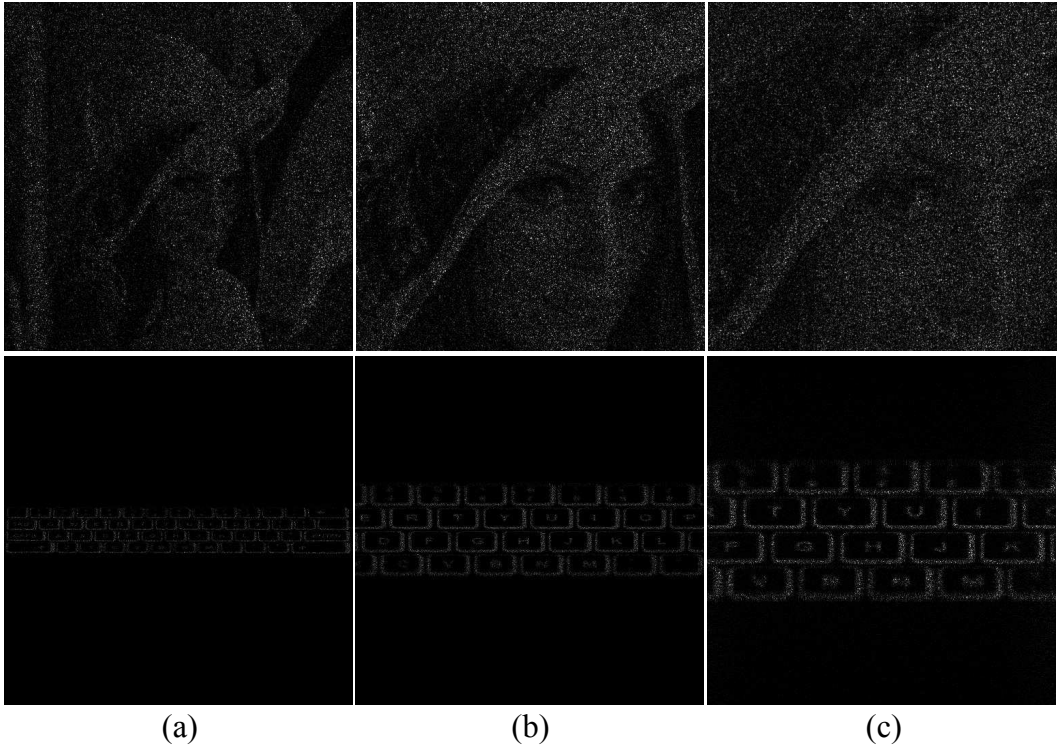}}
\caption{Projected images without optimization from this lensless zoomable holographic projection on tilted planes. (a) $8 \mu$m (b) $16 \mu$m (c) $24 \mu$m.}
\label{fig:reconst_nomulti_notilted}
\end{figure}

\begin{figure}[htb]
\centerline{\includegraphics[width=14cm]{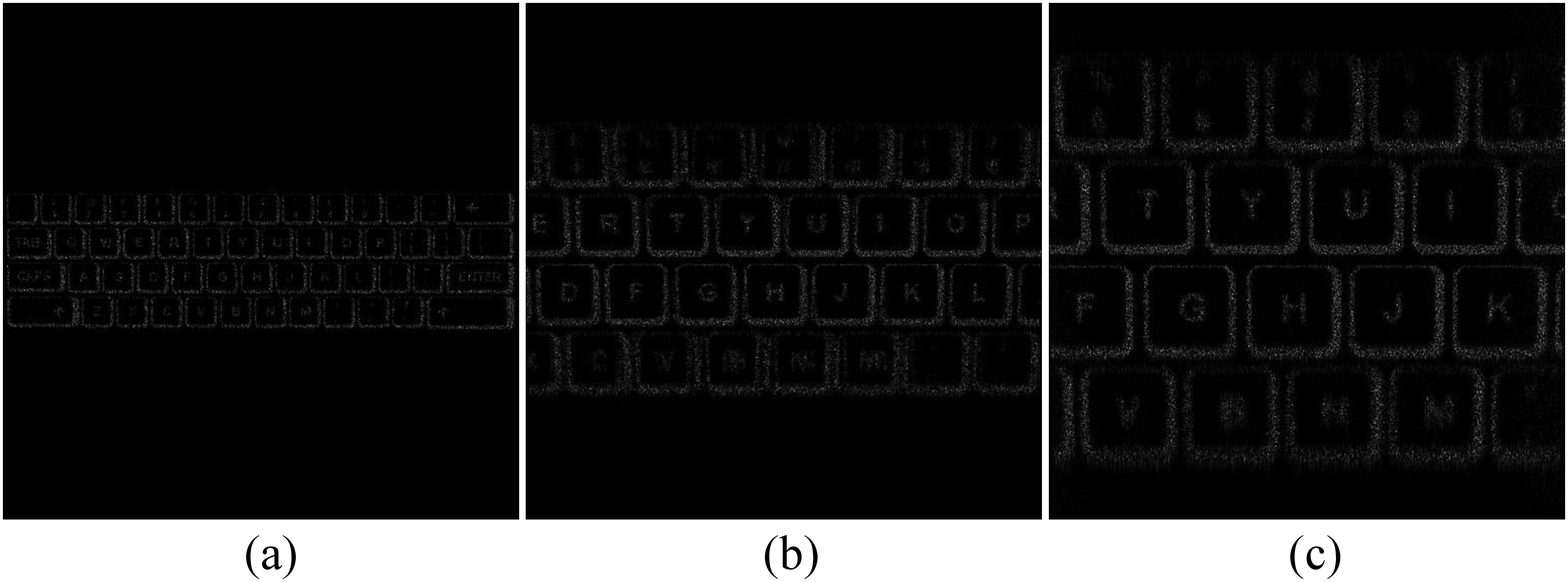}}
\caption{Projected images without optimization from this lensless zoomable holographic projection on tilted planes. (a) $p_1=8 \mu$m (b) $p_1=16 \mu$m (c) $p_1=24 \mu$m..}
\label{fig:reconst_nomulti_tilted}
\end{figure}

Figure \ref{fig:reconst_nomulti_notilted} shows the reconstructed images on the two projection planes when changing the sampling pitch on each image of $8,16,24 \mu$m.
Regarding Projected Image 2, we do not observe it on the tilted plane but observe it in the parallel plane to the hologram.
Therefore, the vertical distortion of Projected Image 2 is observed.
Figure \ref{fig:reconst_nomulti_tilted} shows the reconstructed image on the tilted projection plane.+
We can exactly observe the projected image without the distortion.

As shown in Figs. \ref{fig:reconst_nomulti_notilted} and \ref{fig:reconst_nomulti_tilted}, these reconstructed images are low contrast and contaminated by speckle noise.
In order to improve these problems, iterative algorithms based on GS algorithm \cite{gs} are widely used.
In simulation, the iterative algorithms are well worked; unfortunately, in actual experiments, the iterative algorithms are not because of the difference in the conditions between the simulation and the actual experiment.
Instead of iterative manners, in this paper we use a multi-random phase method \cite{specklegs2} that prepares multiple holograms with different random phases, and then reduces the speckle noise on the reconstructed image by fast switching the holograms, owing to the time averaging effect of human eyes. 

\begin{figure}[htb]
\centerline{\includegraphics[width=14cm]{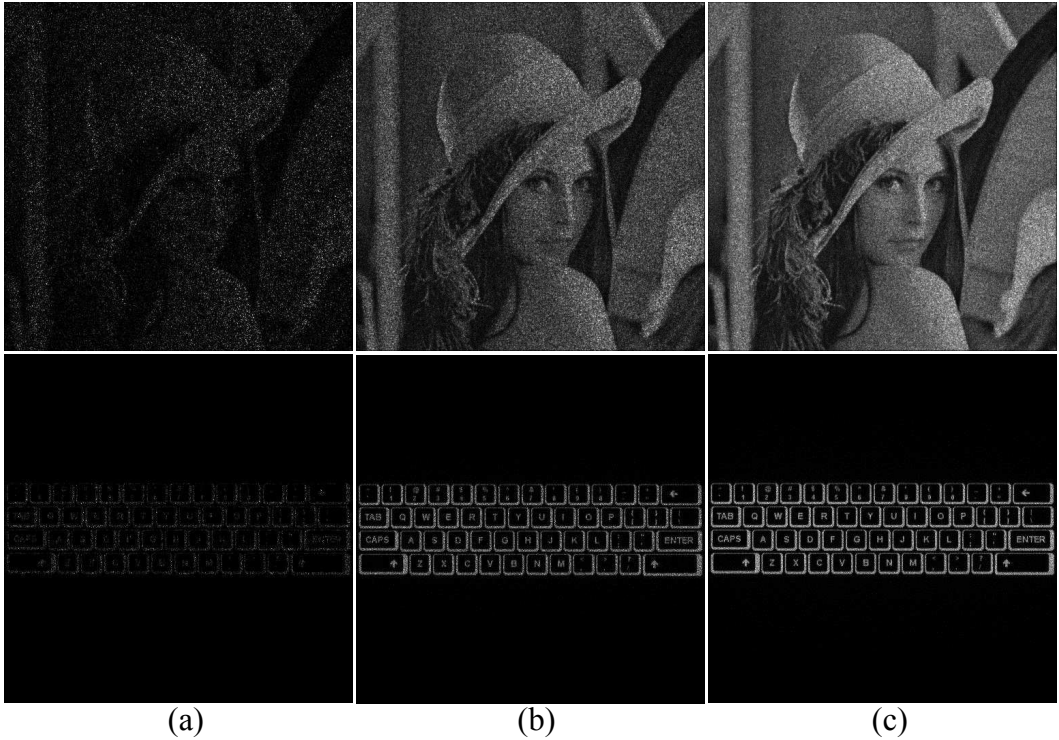}}
\caption{Projected images with speckle reduction from this lensless zoomable holographic projection. (a) one hologram (b) 10 holograms (c) 30 holograms.}
\label{fig:reconst_multi_8um}
\end{figure}

Figure \ref{fig:reconst_multi_8um} shows the reconstructed images when changing the number of temporal superimposing holograms.
The sampling pitches on Images 1 and 2 are the same as that of the hologram.
Figure \ref{fig:reconst_multi_8um} (a) shows the reconstructed images on the projected planes 1 and 2 with a superimposing number of 1.
The reconstructed images are contaminated by speckle noise.
While, Fig.\ref{fig:reconst_multi_8um} (c) shows the reconstructed images with a superimposing number of 30.
The speckle noise is reduced.

\begin{figure}[htb]
\centerline{\includegraphics[width=14cm]{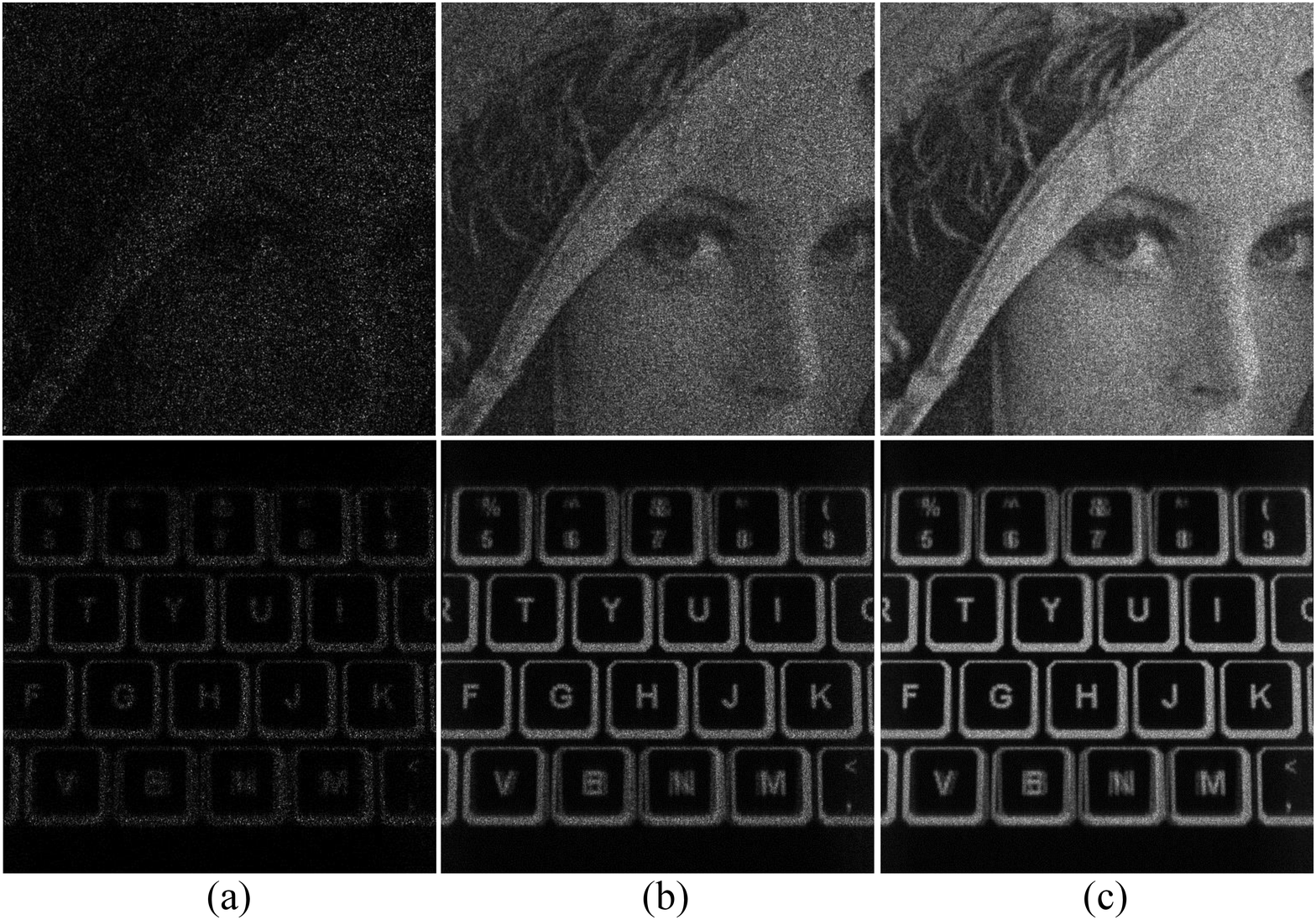}}
\caption{Projected images with speckle reduction from this lensless zoomable holographic projection. (a) one hologram (b) 10 holograms (c) 30 holograms.}
\label{fig:reconst_multi_24um}
\end{figure}

Figure \ref{fig:reconst_multi_24um} shows the reconstructed images when changing the number of temporal superimposing holograms.
The sampling pitches on Images 1 and 2 are three times larger than that of the hologram, that is, these reconstructed images are magnified three-fold.
Figure \ref{fig:reconst_multi_24um} (c) shows the reconstructed images with a superimposing number of 30.
The speckle noise is suppressed.

\begin{figure}[htb]
\centerline{\includegraphics[width=12cm]{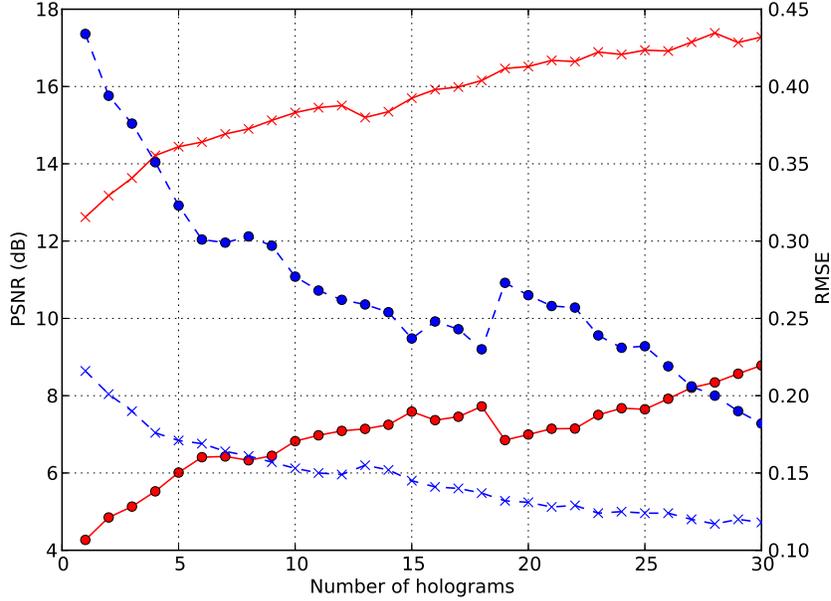}}
\caption{PSNR and RMSE of projected Images 1 and 2 with $p_1=8\mu$m.  The red solid lines with circle and x-mark show the PSNRs of the projected Images 1 and 2, respectively. The blue dashed lines with circle and x-mark show the RMSE of the projected Images 1 and 2, respectively.}
\label{fig:psnr}
\end{figure}

\begin{figure}[htb]
\centerline{\includegraphics[width=12cm]{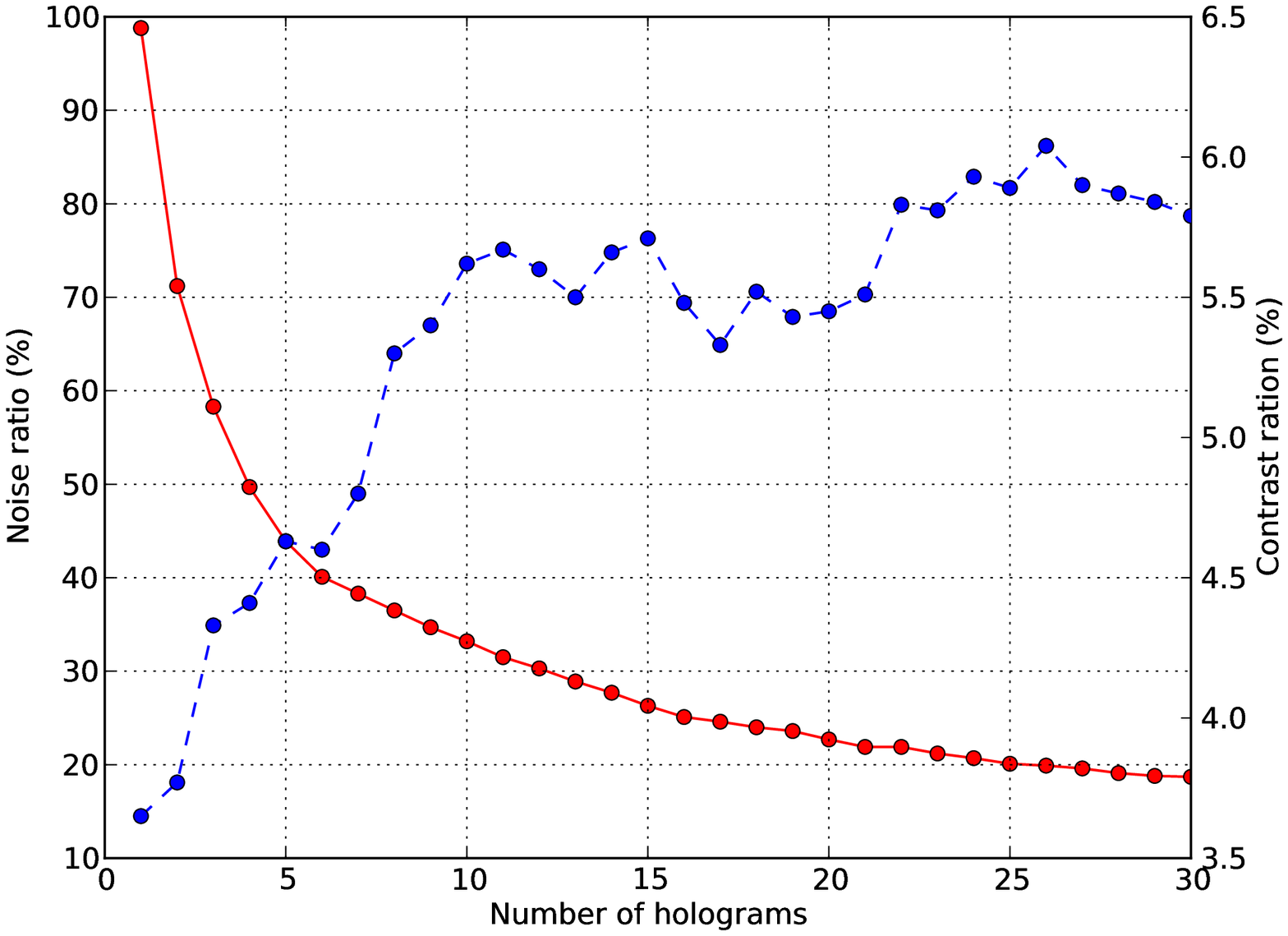}}
\caption{Noise and contrast ratios of projected Images 1 with $p_1=8 \mu$m.  The noise and contrast ratios are the red solid line blue dashed line, respectively.}
\label{fig:noise}
\end{figure}

Figure \ref{fig:psnr} shows the peak signal-to-noise ratio (PSNR) and root mean square error (RMSE) between projected Images 1, 2 and the original images with $p_1=8 \mu$m. 
The red solid lines with circle and x-mark show the PSNRs of the projected Images 1 and 2, respectively.
The blue dashed lines with circle and x-mark show the RMSE of the projected Images 1 and 2, respectively.
As we can see, the PSNR is gradually improved when increasing the number of temporal superimposing holograms.
In addition, we measure noise ratio (i.e. the standard deviation in the intensity in a bright test region divided by the average intensity in the same region) and contrast ratio (i.e. the average intensity in a bright test region divided by the average intensity in a dark test region) on the projected Images 1.
Figure \ref{fig:noise} shows the noise ratio (the red solid line) and contrast ratio (the blue dashed line), respectively.
These metrics are also gradually improved when increasing the number of temporal superimposing holograms.
We used our computational wave optics library, CWO++ \cite{cwo}, in the calculation above.

\section{Conclusion}
This study numerically investigated the feasibility of lensless zoomable holographic multiple projections to tilted planes.
This function was realized by a combination of the scaled diffraction and tilted diffraction calculations.
In addition, we reduced the speckle noise of the projected images on the tilted planes by temporally superimposing holograms.
In our next work, we will demonstrate this lensless zoomable holographic multiple projection in an actual optical setup, and improve the image quality of the projected images by pixel separation method \cite{pixel}.
In this future work, we will use a fast switching spatial light modulator, such as a digital micromirror device (DMD).

\section*{Acknowlegement}
This work is partially supported by JSPS KAKENHI Grant Numbers 25330125 and 25240015, and the Kayamori Foundation of Information Science Advancement and Yazaki Memorial Foundation for Science and Technology.


\end{document}
